\documentclass{acm_proc_article-sp}

\usepackage{graphicx}
\usepackage{amsmath}
\usepackage{amssymb}
\usepackage{tikz}
\usetikzlibrary{decorations.pathmorphing}

\usepackage{hyperref}
\usepackage[colorinlistoftodos, textwidth=3.2cm, shadow]{todonotes}

\begin{document}

\newcommand{\mv}[1]{\ensuremath{\operatorname{\mathit{#1}}}}
\definecolor{dark}{gray}{.6}
\newcommand{\bc}[1]{\textcolor{dark}{#1}}
\newtheorem{lems}{Lemma}
\newtheorem{props}{Proposition}
\newtheorem{thms}{Theorem}
\newtheorem{defs}{Definition}
\newtheorem{obs}{Observation}

\newcommand{\storex}{MRP-Store}

\title{Building global and scalable systems with \\ Atomic Multicast}


\newcommand{\TBS}[1]{\bigskip\noindent\fbox{\parbox{\columnwidth}{\small #1}}\bigskip}
\newcommand{\myparagraph}[1]{\vspace{6pt}\noindent\textbf{#1~}}

\newcommand{\myalign}{-30mm}
\numberofauthors{4} 
\author{
\alignauthor
\hspace{-20mm}Samuel Benz\\
\hspace{-20mm}       \affaddr{University of Lugano}\\
\hspace{-20mm}      \affaddr{Switzerland}\\
\alignauthor
\hspace{\myalign}Parisa Jalili Marandi\\
\hspace{\myalign}       \affaddr{University of Lugano}\\
\hspace{\myalign}       \affaddr{Switzerland}\\
\alignauthor
\hspace{\myalign}Fernando Pedone\\
\hspace{-31mm}       \affaddr{University of Lugano}\\
\hspace{\myalign}       \affaddr{Switzerland}\\
\alignauthor
\hspace{-43mm}Beno\^{i}t Garbinato\\
\hspace{-44mm}       \affaddr{University of Lausanne}\\
\hspace{-44mm}       \affaddr{Switzerland}\\
}

\maketitle

\begin{abstract}

The rise of worldwide Internet-scale services demands large distributed systems. 
Indeed, when handling several millions of users, it is common to operate thousands of servers spread across the globe.
Here, replication plays a central role, as it contributes to improve the user experience by hiding failures and by providing acceptable latency.
In this paper, we claim that atomic multicast, with strong and well-defined properties, is the appropriate abstraction to efficiently design and implement globally scalable distributed systems.
We substantiate our claim with the design of two modern online services atop atomic multicast, a strongly consistent key-value store and a distributed log.
In addition to presenting the design of these services, we experimentally assess their performance in a geographically distributed deployment.

%
%

\end{abstract}

\section{Introduction}
\label{sec:introduction}

In little less that two decades, we have witnessed the explosion of worldwide online services (e.g., search engines, e-commerce, social networks).
These systems typically run on some cloud infrastructure, hosted by datacenters placed around the world.
Moreover, when handling millions of users located everywhere on the planet, it is common for these services to operate thousands of servers scattered across the globe.
A major challenge for such services is to remain available and responsive in spite of server failures and an ever-increasing user base.
Replication plays a key role here, by making it possible to hide failures and to provide acceptable response time.

While replication can potentially lead to highly scalable and available systems, it poses additional challenges. 
Indeed, keeping multiple replicas consistent is a problem that has puzzled system designers for many decades.
Although much progress has been made in the design of consistent replicated systems~\cite{Repl10}, novel application requirements and environment conditions (e.g., very large user base, geographical distribution) continue to defy designers.
Some proposals have responded to these ``new challenges" by weakening the consistency guarantees offered by  services.
Weak consistency is a natural way to handle the complexity of building scalable systems, but it places the burden on the service users, who must cope with non-intuitive service behavior.
Dynamo \cite{DeCandia07}, for instance, overcomes the implications of eventual consistency by letting the users decide about the correct interpretation of the returned data. 
While weak consistency is applicable in some cases, it can be hardly generalized, which helps explain why we observe a recent trend back to strong consistency~(e.g., \cite{aguilera2007sinfonia,balakrishnan2013tango,CDE12,thomson2012calvin}). 
%

Strong consistency entails ordering requests across the system.
Different strategies have been proposed to order requests in a distributed system, which can be divided into two broad categories: those that impose a total order on requests and those that partially order requests.
Many distributed systems today ensure some level of strong consistency by totaling ordering requests using the Paxos algorithm~\cite{Lam98}, or a variation thereof.
For example, Chubby~\cite{B06} is a Paxos-based distributed locking service at the heart of the Google File System (GFS); Ceph~\cite{weil2006ceph} is a distributed file system that relies on Paxos to provide a consistent cluster map to all participants; and Zookeeper~\cite{hunt2010zookeeper} turns a Paxos-like total order protocol into an easy-to-use interface to support group messaging and distributed locking.

In order to scale, services typically partition their state and strive to only order requests that depend on each other, imposing a partial order on requests.
Sinfonia~\cite{aguilera2007sinfonia} and S-DUR~\cite{SPJ12}, for example, build a partial order by using a two-phase commit-like protocol to guarantee that requests spanning common partitions are processed in the same order at each partition.
Spanner~\cite{CDE12} orders requests within partitions using Paxos and across partitions 
using a protocol that computes a request's final timestamp from temporary timestamps proposed by the involved partitions.
%
In this paper, we contend that instead of building a partial order on requests using an ad-hoc protocol intertwined with the application code, services have much to gain from relying on a middleware to partially order requests, analogously to how some libraries provide total order as a service (e.g., \cite{ADM+04}).

Reliably delivering requests in total and partial order has been encapsulated by atomic broadcast and atomic multicast, respectively~\cite{HT93}.
In this paper, we extend Multi-Ring Paxos, a scalable atomic multicast protocol introduced in \cite{MPP2012}, to (a)~cope with large-scale environments and to (b)~allow services to recover from a wide range of failures (e.g., the failures of all replicas).
Addressing these aspects required a redesign of Multi-Ring Paxos and a brand-new library implementation: 
Some large-scale environments (e.g., public datacenters, wide-area networks) do not allow network-level optimizations (e.g., IP-multicast~\cite{MPP2012}) that can significantly boost bandwidth.
Recovering from failures in Multi-Ring Paxos is challenging because it must account for the fact that replicas may not all have the same state.
Thus, a replica cannot recover by installing any other replica's image.

We developed two services based on Multi-Ring Paxos: MRP-Store, a key-value store, and dLog, a distributed log.
These services are at the core of many internet-scale applications.
In both cases, we show in the paper that the challenge of designing and implementing highly available and scalable services can be significantly simplified if these services rely on atomic multicast.
Our performance evaluation assesses the behavior of Multi-Ring Paxos under various conditions and shows that MRP-Store and dLog can scale in different scenarios.
We also illustrate the behavior of MRP-Store when servers recover from failures.

This paper makes the following contributions.
First, we propose an atomic multicast protocol capable of supporting at the same time \emph{scalability} and \emph{strong consistency} in the context of large-scale online services.
Intuitively, Multi-Ring Paxos composes multiple instances of Ring Paxos to provide efficient message ordering.
The Multi-Ring Paxos protocol we describe in the paper does not rely on network-level optimizations (e.g., IP-multicast) and allow services to recover from a wide range of failures.
Second, we show how to design two services, MRP-Store and dLog, atop Multi-Ring Paxos and demonstrate the advantages of our proposed approach.
Third, we detail the implementation of Multi-Ring Paxos, MRP-Store, and dLog.
Finally, we provide a performance assessment of all these components.

The remainder of this paper is structured as follows. 
Section~\ref{sec:model} describes our system model and assumptions.
Section~\ref{sec:whyamcast} explains why system designers must care about atomic multicast as a middleware service.
Sections~\ref{sec:mrpaxos} and~\ref{sec:recovery} present the design of Multi-Ring Paxos and its recovery protocol.
Section~\ref{sec:services} discusses the two services we designed and Section~\ref{prototypes} explains how they were implemented.
Section~\ref{sec:perf} assesses the performance of the components.
Section~\ref{sec:rwork} evaluates the work and Section~\ref{sec:conclusions} concludes this paper.

\section{System model}
\label{sec:model}

We assume a distributed system composed of a set $\Pi = \{ p_1, p_2, ... \}$ of interconnected processes that communicate through the primitives \emph{send}$(p,m)$ and \emph{receive}$(m)$, where $m$ is a message and $p$ is a process.
Processes may fail by crashing and subsequently recover, but do not experience arbitrary behavior (i.e., no Byzantine failures). 
Processes are either \emph{correct} or \emph{faulty}. 
A correct process is eventually operational ``forever" and can reliably exchange messages with other correct processes. 
In practice, ``forever" means long enough for processes to make some progress (e.g., terminate one instance of consensus).
Our protocols ensure safety under both asynchronous and synchronous execution periods. 
To ensure liveness, we assume the system is \emph{partially synchronous}~\cite{DLS88}: it is initially asynchronous and 
eventually becomes synchronous. 
The time when the system becomes synchronous, called the \emph{Global Stabilization Time (GST)}~\cite{DLS88}, is unknown to the processes. 
Before GST, there are no bounds on the time it takes for messages to be transmitted and actions to be executed. 
After GST, such bounds exist but are unknown. 

Atomic multicast is a communication abstraction defined by the primitives \emph{multicast}$(\gamma,m)$ and \emph{deliver}$(m)$, where $m$ is a message and $\gamma$ is a multicast group.
Processes choose from which multicast groups they wish to deliver messages.
If process $p$ chooses to deliver messages multicast to group $\gamma$, we say that $p$ \emph{subscribes to} group $\gamma$.
Let relation $<$ be defined such that $m < m'$ iff there is a process that delivers $m$ before $m'$.
Atomic multicast ensures that 
(i)~if a process delivers $m$, then all correct processes that subscribe to $\gamma$ deliver $m$ \emph{(agreement)}; 
(ii)~if a correct process $p$ multicast $m$ to $\gamma$ then all correct processes that subscribe to $\gamma$ deliver $m$ \emph{(validity)}; and
(iii)~relation $<$ is acyclic \emph{(order)}.
The order property implies that if processes $p$ and $q$ deliver messages $m$ and $m'$, then they deliver them in the same order.
Atomic broadcast is a special case of atomic multicast where there is a single group to which all processes subscribe.

\section{Why Atomic Multicast}
\label{sec:whyamcast}

Two key requirements for current online services are (1)~the immunity to a wide range of failures and (2)~the ability to serve an increasing number of user requests.
The first requirement is usually fulfilled through \emph{replication} within and across datacenters, possibly located in different geographical areas.
%
The second requirement is satisfied through \emph{scalability}, which can be ``horizontal" or ``vertical."
Horizontal scalability (often simply \emph{scalability}) consists in adding more servers to cope with load increases, whereas vertical scalability consists in adding more resources (e.g., processors, disks) to a single server.
Horizontal scalability boils down to partitioning the state of the replicated service and assigning partitions (i.e., so-called shards) to the aforementioned geographically distributed servers.


\myparagraph{Consistency vs. scalability.}
The partition-and-replicate approach raises a challenging concern: How to preserve service consistency in the presence of requests spanning multiple partitions, each partition located in a separate data center, in particular when failures occur?
When addressing this issue, middleware solutions basically differ in how they prioritize \emph{consistency vs.~scalability}, depending on the semantics requirements of the end-user services they support.
That is, while some services choose to relax consistency in favor of scalability and low latency, others choose to tolerate higher latency, possibly sacrificing availability (or at least its perception by end-users), in the interest of service integrity.

\myparagraph{Prioritizing scalability.}
TAO, Facebook's distributed data store~\cite{bronson2013tao}, is an example of a middleware solution that prioritizes scalability over consistency: with TAO, strong consistency is ensured within partitions and a form of eventual consistency is implemented across partitions. 
This implies that concurrent requests accessing multiple partitions may lead to inconsistencies in Facebook's social graph.
To lower potential conflicts, data access patterns can be considered when partitioning data (e.g., entries often accessed together can be located in the same partition). 
Unfortunately, such optimizations are only possible if knowledge about data usage is known a priori, which is often not the case.

Some middleware solutions, such as S-DUR~\cite{SPJ12} and Sinfonia~\cite{aguilera2007sinfonia}, rely on two-phase commit~\cite{bernstein1987concurrency} to provide strong consistency across partitions.
Scatter~\cite{scatter} on the other hand prohibits cross-partition requests and uses a two-phase commit protocol to merge commonly accessed data into the same partition.
A common issue with storage systems that rely on atomic commitment is that requests spanning multiple partitions (e.g., cross-partition transactions) are not totally ordered and can thus invalidate each other, leading to multiple aborts.
For example, assume objects $x$ and $y$ in partitions $p_x$ and $p_y$, respectively, and two transactions $T_1$ and $T_2$ where $T_1$ reads $x$ and updates the value of $y$, and $T_2$ reads $y$ and updates the value of $x$.
If not ordered, both transactions will have to abort to ensure strong consistency (i.e., serializability).

\myparagraph{Prioritizing consistency.}
When it comes to prioritizing consistency, some proposals totally order requests before their execution, as in state-machine replication~\cite{Sch90}, or execute requests first and then totally order the validation of their execution, as in deferred update replication~\cite{PGS02}.
(With state-machine replication requests must execute sequentially; with deferred update replication requests can execute concurrently.)
Coming back to our example of conflicting transactions~$T_1$ and~$T_2$, while approaches based on two-phase commit lead both transactions to abort, with deferred update replication only one transaction aborts~\cite{PGS98}, and with state-machine replication both transactions commit.
Many other solutions based on total order exist, such as Spanner~\cite{CDE12} and Calvin~\cite{thomson2012calvin}.

The Isis toolkit~\cite{birman1990isis} pioneered the use of totally ordered group communication at the middleware level.
With Isis, total order was enforced at two levels: first, a consistent sequence of views listing the replicas considered alive was atomically delivered to each replica; then, messages could be totally ordered within each view, using an atomic broadcast primitive.
In the same vein, many middleware solutions rely on atomic broadcast as basic communication primitive to guarantee total order.

\myparagraph{The best of both worlds.}
We argue that atomic multicast is the right communication abstraction when it comes to combining consistency and scalability.
Indeed, atomic broadcast implies that all replicas are in the same group and must thus receive each and every request, regardless its actual content, which makes atomic broadcast an inefficient communication primitive when data is partitioned and possibly spread across datacenters.
With atomic multicast, on the contrary, each request is only sent to the replicas involved in the request, which is more efficient when data is partitioned and possibly distributed across datacenters.
Compared to solutions that rely on atomic broadcast to ensure consistency within each partition and an ad hoc protocol to handle cross-partition requests, atomic multicast is more advantageous in that requests are ordered both within and across partitions.

Not only do we advocate atomic multicast as basic communication primitive to build middleware services, we also believe that the traditional group addressing semantics should be replaced with one that better fits the context of large-scale Internet services.
With traditional atomic multicast primitives (e.g., \cite{DGF00,GS01b,rodrigues1998scalatom, schiper2008inherent,schiper2010p}), a client can address multiple \emph{non-intersecting} groups of servers, where each server can only belong to a single group.
Rather, we argue that clients should address one group per multicast and each server should be able to subscribe to any group it is interested in, i.e., any replication group corresponding to the shards the server is currently replicating, similarly to what IP~multicast supports.
As we shall see in Section~\ref{sec:mrpaxos}, this somehow ``inverted'' group addressing semantics allow us to implement a scalable atomic multicast protocol.

\myparagraph{Recovering from failures.}
The ability to safely recover after a failure is an essential aspect of the failure immunity requirement on large-scale middleware services. 
Furthermore, fast crash recovery is of practical importance when in-memory data structures are used to significantly decrease latency.
Yet similarly to what is done to ensure cross-partition consistency, existing middleware solutions tend to deal with recovery issues in an ad~hoc manner, directly at the service level, rather than factor out the solution to recovery issues in the underlying communication layer.
A different approach consists in relying on atomic multicast to orchestrate checkpointing and coordinate checkpoints with the trimming of the logs used by the ordering protocol.
This is particularly important in the context of atomic multicast since recovery in partitioned systems is considerably more complex than recovery in single partition systems (see Section~\ref{sec:recovery}).

\begin{figure}[ht]
 \centering
 \includegraphics[scale=0.6]{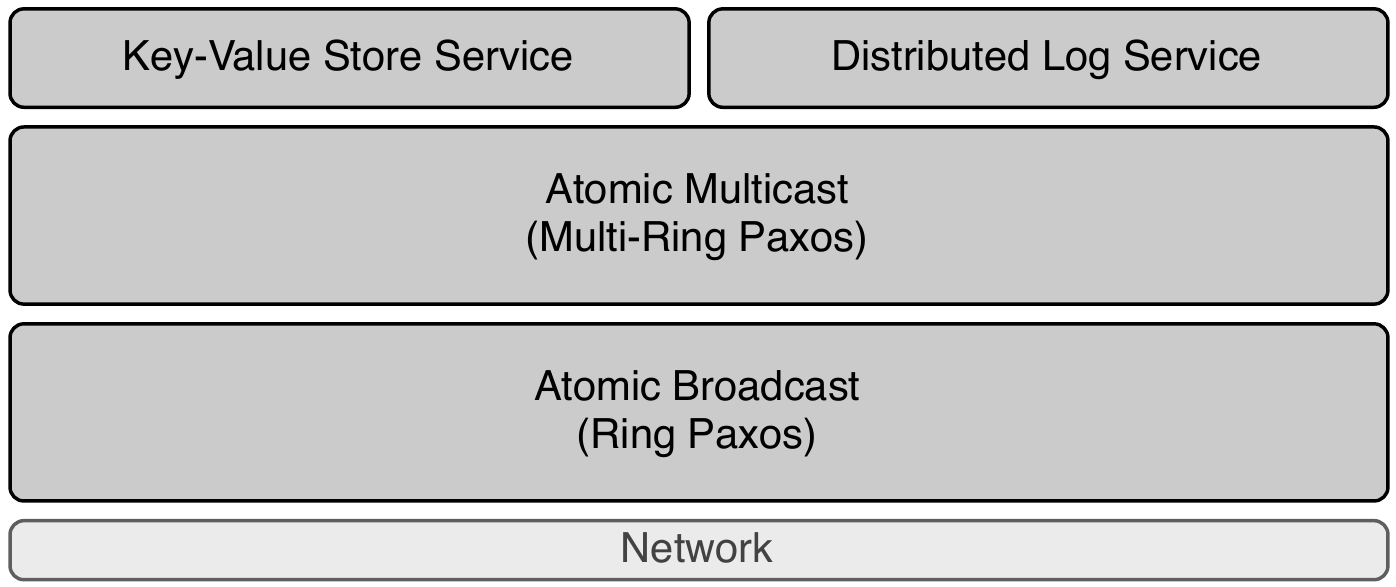}
 \caption{Architecture overview.}
 \label{fig:architecture}
\end{figure}

\myparagraph{Architecture overview.}
Figure~\ref{fig:architecture} presents an overview of our middleware solution based on atomic multicast, implemented by Multi-Ring Paxos.
Online services can build on atomic multicast's ordering and recovery properties, as described in the next two sections. 
As suggested by this figure, atomic multicast naturally supports state partitioning, an important characteristic of scalable services, and no ad~hoc protocol is needed to handle coordination among partitions.

\section{Multi-Ring Paxos}
\label{sec:mrpaxos}

Intuitively, Multi-Ring~Paxos turns an atomic broadcast protocol based on Ring Paxos into an atomic multicast protocol.
That is, Multi-Ring~Paxos is implemented as a collection of coordinated Ring Paxos instances, or rings for short, such that a distinct multicast group is assigned to each ring.
Each ring in turn relies on a sequence of consensus instances, implemented as an optimized version of Paxos.

Multi-Ring Paxos was introduced in~\cite{MPP2012}.
In this section, we recall how Multi-Ring Paxos works and describe a variation of Ring Paxos that does not rely on network-level optimizations (e.g., IP-multicast) to achieve high throughput.
In the next section, we introduce Multi-Ring Paxos's recovery.

\begin{figure*}[ht]
 \centering
 \includegraphics[scale=0.8]{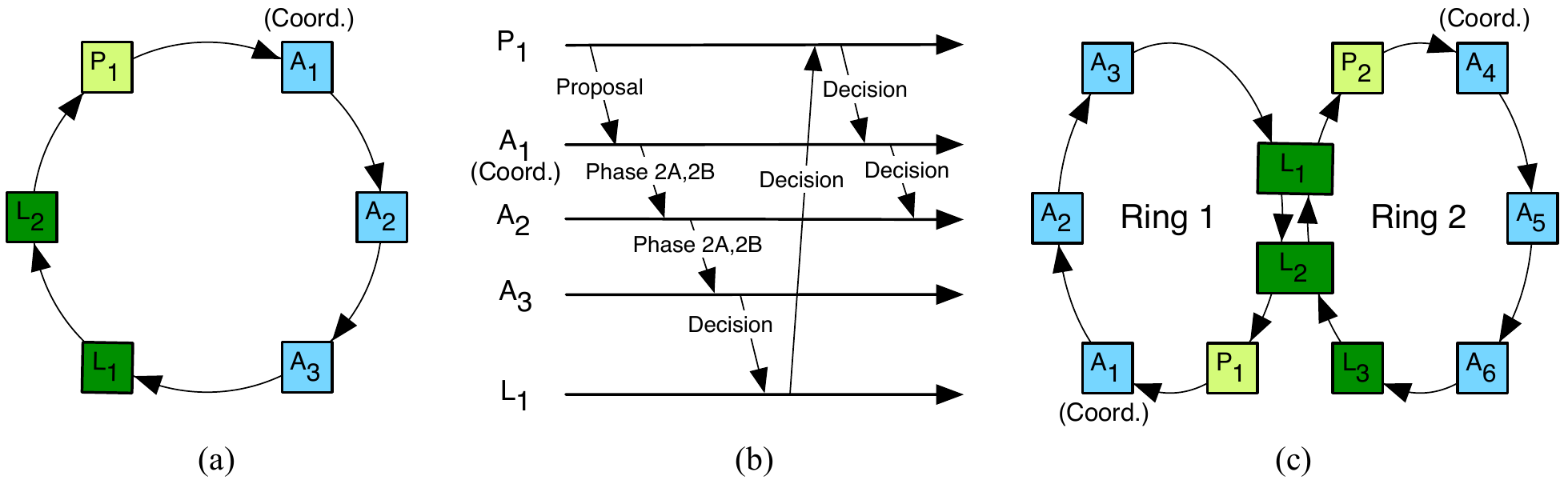}
 \caption{(a)~The various process roles in Ring Paxos disposed in one logical ring; (b)~an execution of a single instance of Ring Paxos; and (c)~a configuration of Multi-Ring Paxos involving two rings (learners $L_1$ and $L_2$ deliver messages from Rings 1 and 2, and leaner $L_3$ delivers messages from Ring 2 only).}
 \label{fig:rpaxos}
\end{figure*}

\myparagraph{Ring Paxos.}
Similarly to Paxos, Ring Paxos differentiates processes as \emph{proposers}, \emph{acceptors}, and \emph{learners}, where one of the acceptors is elected as the \emph{coordinator}. 
All processes in Ring Paxos communicate through a unidirectional ring overlay, as illustrated in Figure~\ref{fig:rpaxos}~(a). 
Using a ring topology for communication enables a balanced use of networking resources and results in high performance. 

Figure~\ref{fig:rpaxos}~(b)  illustrates the operations of an optimized Paxos, where Phase~1 is pre-executed for a collection of instances.
When a proposer proposes a value (i.e., the value is atomically broadcast), the value circulates along the ring until it reaches the coordinator. 
The coordinator proposes the value in a Phase~2A message and forwards it to its successor in the ring together with its own vote, that is, a Phase~2B message. 
If an acceptor receives a Phase~2A/2B message and agrees to vote for the proposed value, the acceptor updates Phase~2B with its vote and sends the modified Phase~2A/2B message to the next process in the ring.
If a non-acceptor receives a Phase 2A/2B message, it simply forwards the message as is to its successor. 
When the last acceptor in the ring receives a majority of votes for a value in a Phase~2B message, it replaces the Phase~2B message by a decision message and forwards the outcome to its successor. 
Values and decisions stop circulating in the ring when all processes in the ring have received them. 
A process learns a value once it receives the value and the decision that the value can be learned (i.e., the value is then delivered).
To optimize network and CPU usage, different types of messages for several consensus instances (e.g., decision, Phase~2A/2B) are often grouped into bigger packets before being forwarded. 
Ring Paxos is oblivious to the relative position of processes in the ring.
Ring configuration and coordinator's election are handled with a coordination system, such as~Zookeeper.

\myparagraph{Multi-Ring Paxos.}
With Multi-Ring Paxos, each Learner can subscribe to as many rings as it wants and participates in coordinating multiple instances of Ring Paxos for those rings. 
In Figure~\ref{fig:rpaxos}~(c), we picture a deployment of Multi-Ring Paxos with two rings and three learners, where learners L1 and L2 subscribe to rings 1 and 2, and learner L3 subscribes only to ring 2. 
The coordination between groups relies on two techniques, \emph{deterministic merge} and \emph{rate leveling}, controlled with three parameters: $M$, $\Delta$, and $\lambda$. 

Initially, a proposer multicasts a value to group $\gamma$ by proposing the value to the coordinator responsible for $\gamma$.
Then, Learners use a \emph{deterministic merge} strategy to guarantee atomic multicast's ordered delivery property:
Learners deliver messages from rings they subscribe to in round-robin, following the order given by the ring identifier.
More precisely, a learner delivers messages decided in $M$ consensus instances from the first ring, then delivers messages decided in $M$ consensus instances from the second ring, and so on and then starts again with the next $M$ consensus instances from the first ring.

Since multicast groups may not be subject to the same load, with the deterministic merge strategy described above, replicas would deliver messages at the speed of the slowest multicast group, i.e., the group taking the longest time to complete $M$~consensus instances.
To counter the effects of unbalanced load, Multi-Ring Paxos uses a \emph{rate leveling} strategy whereby the coordinators of slow rings periodically propose to skip consensus instances.
That is, at regular $\Delta$ intervals, a coordinator compares the number of messages proposed in the interval with the maximum expected rate $\lambda$ for the group and proposes enough skip instances to reach the maximum rate.
To skip an instance, the coordinator proposes null values in Phase~2A messages. 
For performance, the coordinator can propose to skip several consensus instances in a single message.
\section{Recovery}
\label{sec:recovery}

For a middleware relying on Multi-Ring Paxos to be complete and usable, processes must be able to recover from failures.
More precisely, recovery should allow processes to (a)~restart their execution after failures and (b)~limit the amount of information needed for restart.
Multi-Ring Paxos's recovery builds on Ring Paxos's recovery.
In the following, we first describe recovery in Ring Paxos (Section~\ref{ssec:recrpaxos}) and then detail the subtleties involving recovery in Multi-Ring Paxos (Section~\ref{ssec:recmrpaxos}).

\subsection{Recovery in Ring Paxos}
\label{ssec:recrpaxos}

The mechanism used by a process to recover from a failure in Ring Paxos depends on the role played by the process.
In a typical deployment of Ring Paxos (e.g., state-machine replication~\cite{Lam78,Sch90}), clients propose commands and replicas deliver and execute those commands in the same total order before responding to the clients.
In this case, clients act as proposers and replicas as learners, while acceptors ensure ordered delivery of messages.
In the following, we focus the discussion on the recovery of acceptors and replicas.
Recovering clients is comparatively an easier task.

\myparagraph{Acceptor Recovery.}
Acceptors need information related to past consensus instances in order to serve retransmission requests from recovering replicas.
So, before responding to a coordinator's request with a Phase~1B or Phase~2B message, an acceptor must log its response onto stable storage.
This ensures that upon recovering from a failure, the acceptor can retrieve data related to consensus instances it participated in before the failure.
In principle, an acceptor must keep data for every consensus instance in which it participated.
In practice, it can coordinate with replicas to trim its log, that is, to delete data about old consensus instances.

\myparagraph{Replica Recovery.}
When a replica resumes execution after a failure, it must build a state that is consistent with the state of the replicas that did not crash.
For this reason, each replica periodically checkpoints its state onto stable storage.
Then, upon resuming from a failure, the replica can read and install its last stored checkpoint and contact the acceptors to recover the commands missing from this checkpoint, i.e., the commands executed after the replica's last checkpoint.

\myparagraph{Optimizations.}
The above recovery procedure is optimized as follows.
If the last checkpointed state of a recovering replica is ``too old",\footnote{That is, it would require the processing of too many missing commands in order to build an up-to-date consistent state.} the replica builds an updated state by retrieving the latest checkpoint from an operational replica.
This optimization reduces the number of commands that must be recovered from the acceptors, at the cost of transferring the complete state from a remote replica.

\subsection{Recovery in Multi-Ring Paxos}
\label{ssec:recmrpaxos}

Recovery in Multi-Ring Paxos is more elaborate than in Ring Paxos.
This happens because in Multi-Ring Paxos replicas may deliver messages from different multicast groups and thus evolve through different sequences of states.
We call the set of replicas that deliver messages from the same set of multicast groups a \emph{partition}.
Replicas in the same partition evolve through the same sequence of states.
Therefore, in Multi-Ring Paxos, a recovering replica can only recover a remote checkpoint, to build an updated state, from another replica in the same partition.


As in Ring Paxos, replicas periodically checkpoint their state.
Because a replica $p$'s state may depend on commands delivered from multiple multicast groups, however, $p$'s checkpoint in Multi-Ring Paxos is identified by a tuple $k_p$ of consensus instances, with one entry in the tuple per multicast group.
A checkpoint identified by tuple $k_p$ reflects commands decided in consensus instances up to $k[x]_p$, for each multicast group $x$ that $p$ subscribed to.
Since entries in $k_p$ are ordered by group identifier and replicas deliver messages from groups they subscribe to in round-robin, in the order given by the group identifier, predicate \ref{eq:cond} holds for any state checkpointed by replica~$p$ involving multicast groups $x$ and $y$:
\begin{equation}
\label{eq:cond}
x < y \Rightarrow k[x]_p \geq k[y]_p
\end{equation}
Note that Predicate~\ref{eq:cond} establishes a total order on checkpoints taken by replicas in the same partition.

Periodically, the coordinator of a multicast group $x$ asks replicas that subscribe to $x$ for the highest consensus instance that acceptors in the corresponding ring can use to safely trim their log.
Every replica $p$ replies with its highest safe instance $k[x]_p$ to the coordinator, reflecting the fact that the replica has checkpointed a state containing the effects of commands decided up to instance $k[x]_p$.
The coordinator waits for a quorum $Q_T$ of answers from the replicas, computes the lowest instance number $K[x]_T$ out of the values received in $Q_T$ and sends $K[x]_T$ to all acceptors.
That is, we have that the following predicate holds for~$K[x]_T$:
%
\begin{equation}
\label{eq:1}
\forall p \in Q_T: K[x]_T \leq k[x]_p
\end{equation}
Upon receiving the coordinator's message, each acceptor can then trim its log, removing data about all consensus instances up to instance $K[x]_T$.

A recovering replica contacts replicas in the same partition and waits for responses from a recovery quorum $Q_R$.
Each replica $q$ responds with the identifier $k_q$ of its most up-to-date checkpoint, containing commands up to consensus instances in $k_q$.
The recovering replica selects the replica with the most up-to-date checkpoint available in $Q_R$, identified by tuple $K_R$ such that:
%
\begin{equation}
\label{eq:2}
\forall q \in Q_R: k_q \leq K_R
\end{equation}
If $Q_T$ and $Q_R$ intersect, then by choosing the most up-to-date checkpoint in $Q_R$, identified by $K_R$, the recovering replica can retrieve any consensus instances missing in the selected checkpoint since such instances have not been removed by the acceptors yet.

Indeed, since $Q_T$ and $Q_R$ intersect, there is at least one replica $r$ in both quorums.
For each multicast group $x$ in the partition, from Predicates~\ref{eq:cond} and~\ref{eq:2}, we have $k[x]_r \leq K_R[x]$.
Since $r$ is in $Q_T$, from Predicate~\ref{eq:1}, we have $K_T[x] \leq k[x]_r$ and therefore:
\begin{equation}
\label{eq:3}
K_T \leq k_r \leq K_R
\end{equation}
which then results in:
\begin{equation}
\label{eq:4}
K_T \leq K_R
\end{equation}
Predicate~\ref{eq:4} implies that for every multicast group $x$ in the most up-to-date checkpoint in $Q_R$, the acceptors involved in $x$ have trimmed consensus instances at most equal to the ones reflected in the checkpoint.
Thus, a recovering replica will be able to retrieve any instances decided after the checkpoint was taken.

\section{Services}
\label{sec:services}

We have used two services, a key-value store and a distributed log, to illustrate the capabilities of Multi-Ring Paxos.
In this section we briefly discuss these services. 

\subsection{\storex}

\storex\ implements a key-value store service where keys are strings and values are byte arrays of arbitrary size. 
The database is divided into $l$ partitions $P_0, P_1, ..., P_l$ such that each partition $P_i$ is responsible for a subset of keys in the key space. 
Applications can decide whether the data is hash- or range-partitioned~\cite{OV99}, and clients must know the partitioning scheme. 
The service is accessed through primitives to read, update, insert, and delete an entry (see Table~\ref{tbl:reqs}). 
Additionally we provide a range scan command to retrieve entries whose keys are within a given interval. 

\begin{table}[htdp]
\centering
\begin{tabular}{|l|l|} \hline
Operation	& Description \\ \hline
read$(k)$	& return the value of entry $k$, if existent \\
scan$(k, k')$	& return all entries within range $k..k'$ \\
update$(k,v)$	& update entry $k$ with value $v$, if existent \\
insert$(k,v)$	& insert tuple $(k,v)$ in the database \\
delete$(k)$	& delete entry $k$ from the database \\ 
\hline
\end{tabular}
\caption{\storex\ operations.}
\label{tbl:reqs}
\end{table}

\storex\ replicates each partition using the state-machine replication approach~\cite{Lam98}, implemented with Multi-Ring Paxos. 
A request to read, update, insert, or delete entry $k$ is multicast to the partition where $k$ belongs; a scan request is 
multicast to all partitions that may possibly store an entry within the provided range, if data is range-partitioned, or to all partitions, if data is hash-partitioned.

\storex\ ensures sequential consistency~\cite{AW2004}, that is, there is a way to serialize client operations in any execution such that: 
(1)~it respects the semantics of the objects, as determined in their sequential specifications and 
(2)~it respects the order of non-overlapping operations submitted by the same client.
Atomic multicast prevents cycles in the execution of multi-partitions operations, which would result in non-serializable executions.

\subsection{dLog}

DLog is a distributed shared log that allows multiple concurrent writers to append data to one or multiple logs atomically (see Table~\ref{tbl:dlogiface}). 
Append and multi-append commands return the position of the log at which the data was stored. 
There are also commands to read from a position in a log and to trim a log at a certain position.
Like \storex, dLog uses state-machine replication implemented with Multi-Ring Paxos. 
Commands to append, read, and trim are multicast to the log they address and multi-append commands are multicast to all logs involved.
A dLog server holds the most recent appends in-memory and can be configured to write data asynchronously or synchronously to disk. 


\begin{table}[htdp]
\centering
\begin{tabular}{|l|l|} \hline
Operation	& Description \\ \hline
append$(l, v)$	& append $v$ to log $l$, return position $p$ \\
multi-append$(\mathcal{L},v)$	& append value $v$ to logs in $\mathcal{L}$ \\
read$(l,p)$	& return value $v$ at position $p$ in log $l$ \\
trim$(l,p)$		& trim log $l$ up to position $p$ \\
\hline
\end{tabular}
\caption{dLog operations.}
\label{tbl:dlogiface}
\end{table}

\section{Implementation}
\label{prototypes}

In this section, we discuss important aspects about the implementations 
of Multi-Ring Paxos and the services we built on top of it. 

\subsection{Multi-Ring Paxos}

Multi-Ring Paxos is implemented mostly in Java, with a few parts in C.
All the processes in Multi-Ring Paxos, independent of their roles, are multi-threaded. 
Threads communicate through Java's standard queues. 
A learner has dedicated threads per each ring it subscribes to. 
Another thread then deterministically merges the queues of these threads. 
Acceptors, when using in-memory storage, have access to pre-allocated buffers with 15000 slots, each slot of size 32 Kbytes. 
Disk writes are implemented using the Java version of Berkeley DB. 
All communication within Multi-Ring Paxos is based on TCP. 
Automatic ring management and configuration management is handled by Zookeeper. 
Applications can use Multi-Ring Paxos by including it as a library or by running it standalone. 
In standalone mode, applications can communicate using a Thrift API.\footnote{http://thrift.apache.org/}
Multi-Ring Paxos is publicly available for download.\footnote{https://github.com/sambenz/URingPaxos}

\subsection{MRP-Store}

In our prototype, clients connect to proposers through Thrift and replicas implement the learner interface. 
The partitioning schema is stored in Zookeeper and accessible to all processes. 
Clients determine an entry's location using the partitioning information and send the command to a proposer of the corresponding ring. 
Clients may batch small commands, grouped by partition, up to 32 Kbytes. 
Replicas reply to clients with the response of a command using UDP. 
Clients wait for the first response from a replica in single-partition commands or for at least one response from every partition in scan operations.

Database entries are stored in an in-memory tree at every replica. 
Replicas comply with Multi-Ring Paxos's recovery strategy (see Section~\ref{ssec:recmrpaxos}) by periodically taking checkpoints of the in-memory structure and writing them synchronously to disk. 
After a majority of replicas have written their state to stable storage, Paxos acceptors are allowed to trim their logs. 
A recovering replica will contact a majority of other replicas and download the most recent remote checkpoint.

\subsection{dLog}

Similarly to MRP-Store, dLog clients submit commands to replicas using Thrift.
Multiple commands from one client can be grouped in batches of up to 32 Kbytes. 
Replicas implement the learner's interface to deliver commands.
%
Replicas append the most recent writes to an in-memory cache of 200 Mbytes and write all data asynchronously to disk.
Results from the execution of commands are sent back to clients through UDP. 
%
A trim command flushes the cache up to the trim position and creates a new log file on disk.

\section{Performance evaluation}
\label{sec:perf}

In this section, we experimentally assess various aspects of the performance of our proposed systems: 
\vspace{-4mm}\begin{itemize}
\item We establish a baseline performance for Multi-Ring Paxos, MRP-Store, and dLog. 
\item We measure vertical and horizontal scalability of MRP-Store and dLog in a datacenter and across datacenters. 
\item We evaluate the impact of recovery on performance.
\end{itemize}

\subsection{Hardware setup}

All the ``local experiments" (i.e., within a datacenter) were performed in a cluster of 4 servers equipped with 32-core 2.6 GHz Xeon CPUs and 128 GB of main memory. 
These servers were interconnected through a 48 port 10 Gbps switch with round trip time of 0.1 millisecond. 
In all the experiments, clients and servers were deployed on separate machines. 
For persistency we use solid-state disks (SSDs) with 240 GB and 5 7200-RPM harddisks with 4 TB each. 
Each machine was equipped with 2 NICs of 10 Gbps capacity. 
The ``global experiments" (i.e., across datacenters) were performed on Amazon EC2 with large instances. 
Each large instance server was equipped with 7.5 GB of main memory and a 32 GB local SSD.


\subsection{Experimental setup}

Within a datacenter, Multi-Ring Paxos was initialized as follows: $M=$1, $\Delta=$ 5 millisecond, and $\lambda=$ 9000. Across datacenters, the following configuration was used: $M=$1, $\Delta=$ 20 millisecond, and $\lambda=$ 2000. 
We keep machines approximately synchronized by running the NTP service before the experiments. 
We used Berkeley DB version JE 5.0.58 as persistent storage. 
Unless stated otherwise, acceptors used asynchronous disk writes. 
When in synchronous mode, batching was disabled, that is, instances were written to disk one by one. 
Each experiment is performed for a duration of at least 100 seconds. 

\subsection{Baseline performance}

In this section, we evaluate the performance of a single multicast group in Multi-Ring Paxos with a ``dummy service" (i.e., commands do not execute any operations) under varying request sizes and storage modes.
We also compare the performance of MRP-Store and dLog to existing services with similar functionality.

\subsubsection{Multi-Ring Paxos}

\textbf{Setup.} 
In this experiment there is one ring with three processes, all of which are proposers, acceptors, and learners, and one of the acceptors is the coordinator. 
Proposers have 10 threads, each one submiting requests whose size varies between 512 bytes and 32 Kbytes. 
Batching is disabled in the ring. 
We consider five different storage modes: in-memory, synchronous and asynchronous disk writes using solid-state disks and harddisks.

\begin{figure*}[t]
  \begin{center}
    \begin{tabular}{c}      
      \includegraphics[width=\textwidth]{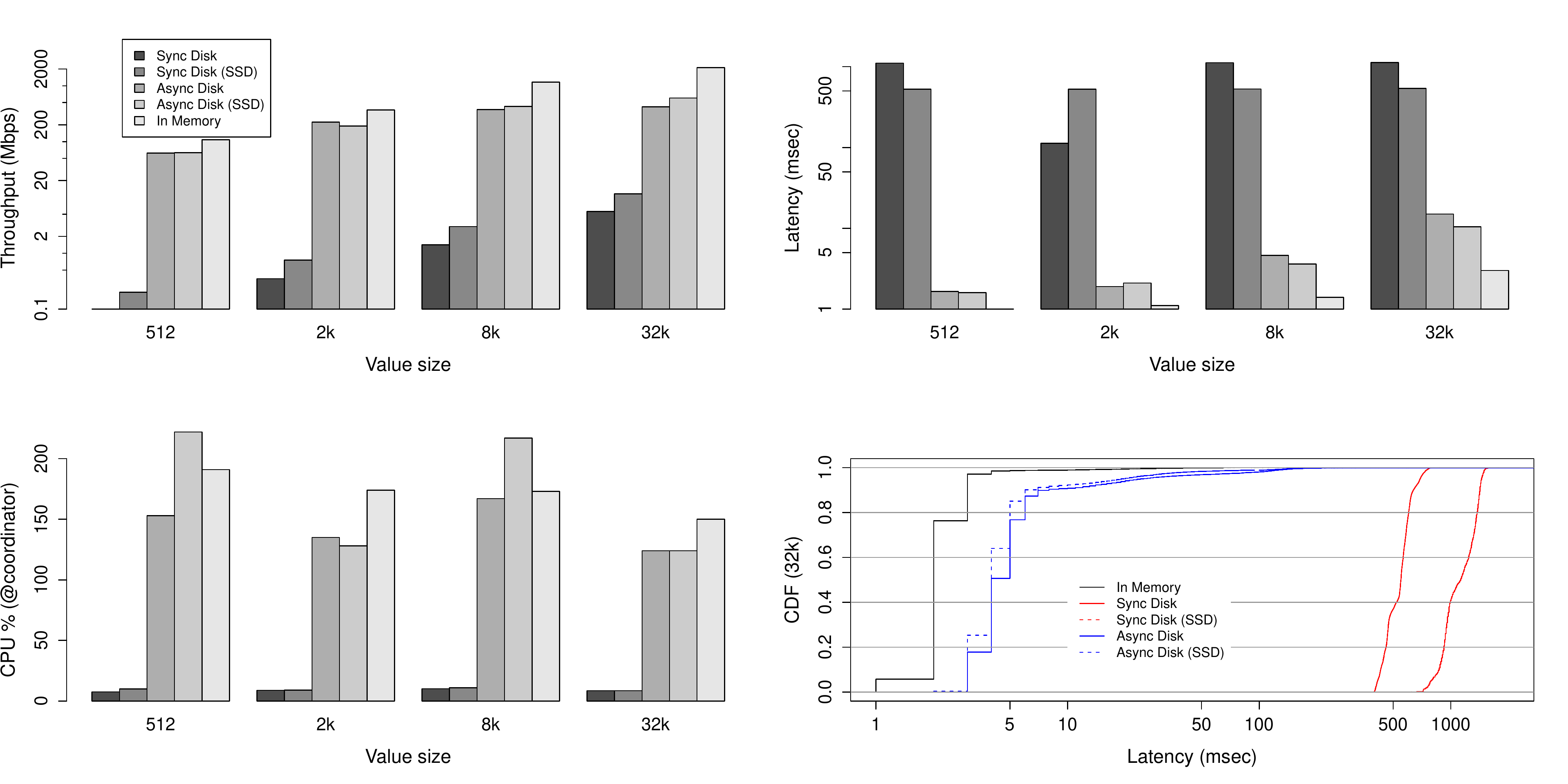} 
    \end{tabular}
\caption{Multi-Ring Paxos with different storage modes and request sizes. Four metrics are measured: throughput in mega bits per second (top-left graph), average latency in milliseconds (top-right graph), CPU utilization at coordinator (bottom-left graph), and CDF for the latency when requests are 32 KBytes (bottom-right graph). The y-axis for throughput and latency is in log scale.}
    \label{fig:ringpaxos}
  \end{center}
\end{figure*}

\textbf{Results.} As seen in the top-left graph of Figure~\ref{fig:ringpaxos}, 
regardless the storage mode, throughput increases as the request size increases. 
With synchronous disk writes, the throughput is limited by the disk's performance. 
With in-memory storage mode, the throughput is limited by the coordinator's CPU (bottom-left graph). 
The coordinator's CPU usage is the highest in asynchronous mode. 
This is due to Java's parallel garbage collection. 
For in-memory storage, we allocate memory outside of Java's heap and therefore performance is not affected by Java's garbage collection. 
%
The bottom-right graph of Figure~\ref{fig:ringpaxos} shows the CDF of latency for 32 Kbyte values. 
In synchronous disk write mode, more than 90\% of requests take less than 10 milliseconds. 

\subsubsection{MRP-Store}

\textbf{Setup.} 
In this experiment, we use Yahoo! Cloud Serving Benchmark (YCSB)~\cite{cooper2010benchmarking} to compare the performance of MRP-Store against Apache's Cassandra and a single MySQL instance. 
These systems provide different consistency guarantees, and by comparing them we can highlight the performance implications of each guarantee. 
In the experiments with MRP-Store, we use three partitions, where participants in a partition subscribe to a ring local to the partition.
Each ring is deployed with three acceptors, all of which write asynchronously to disk.
We test configurations of MRP-Store where replicas in the partitions subscribe to a common global ring and where there is no global ring coordinating the replicas (in the graph, ``independent rings").
All the rings are co-located on three machines and clients run on a separate machine.
In the experiments with Cassandra, we initiate three partitions with replication factor three. 
MySQL is deployed on a single server. 
In all cases, the database is initialized with 1 GByte of data.

\begin{figure}[t]
  \begin{center}
    \begin{tabular}{c}      
      \includegraphics[width=\columnwidth]{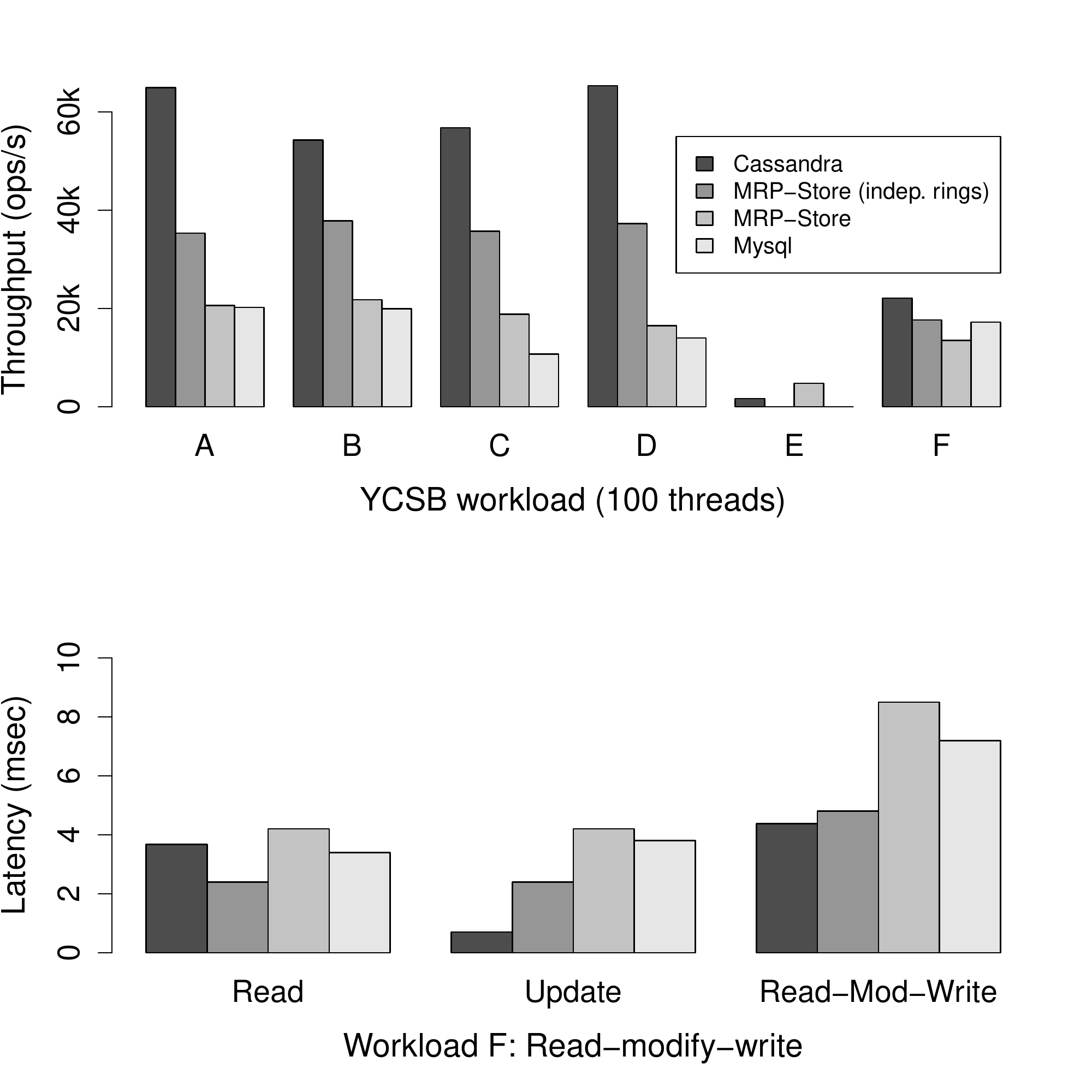} 
    \end{tabular}
\caption{Performance of Apache's Cassandra, two configurations of MRP-Store, and MySQL, under Yahoo! cloud serving benchmark (YCSB). The graphs show throughput in operations per second (top) and average latency in msecs (bottom).}
    \label{fig:cassandra}
  \end{center}
\end{figure}

\textbf{Results.} 
With the exception of Workload E, composed of 95\% of small range scans and 5\% of inserts, Cassandra is consistently more efficient than the other systems since it does not impose any ordering on requests (see Figure~\ref{fig:cassandra}).
Ordering requests within partitions only (i.e., independent rings) is cheeper than ordering requests within and across the system.
This happens because with independent rings, each ring can proceed at its own pace, regardless the load in the other rings. 
To a certain extent, this can be understood as the cost of ensuring stronger levels of consistency.
In our settings, MRP-Store compares similarly to MySQL.
As we show in the following sections, MRP-Store can scale with additional partitions while keeping the same ordering guarantees, something that is not possible with MySQL.


\subsubsection{dLog}
\textbf{Setup.} In this experiment, we compare the performance of our dLog service to Apache's Bookkeeper. 
Both systems implement a distributed log with strong consistency guaranties. 
All requests are written to disk synchronously. 
The dLog service uses two rings with three acceptors per ring. 
dLog learners subscribe to both rings and are co-located with the acceptors. 
Bookkeeper uses an ensemble of the same three nodes. A multithreaded client runs on a different machine and sends append requests of 1 KBytes. 

\begin{figure}
  \begin{center}
    \begin{tabular}{c}      
      \includegraphics[width=\columnwidth]{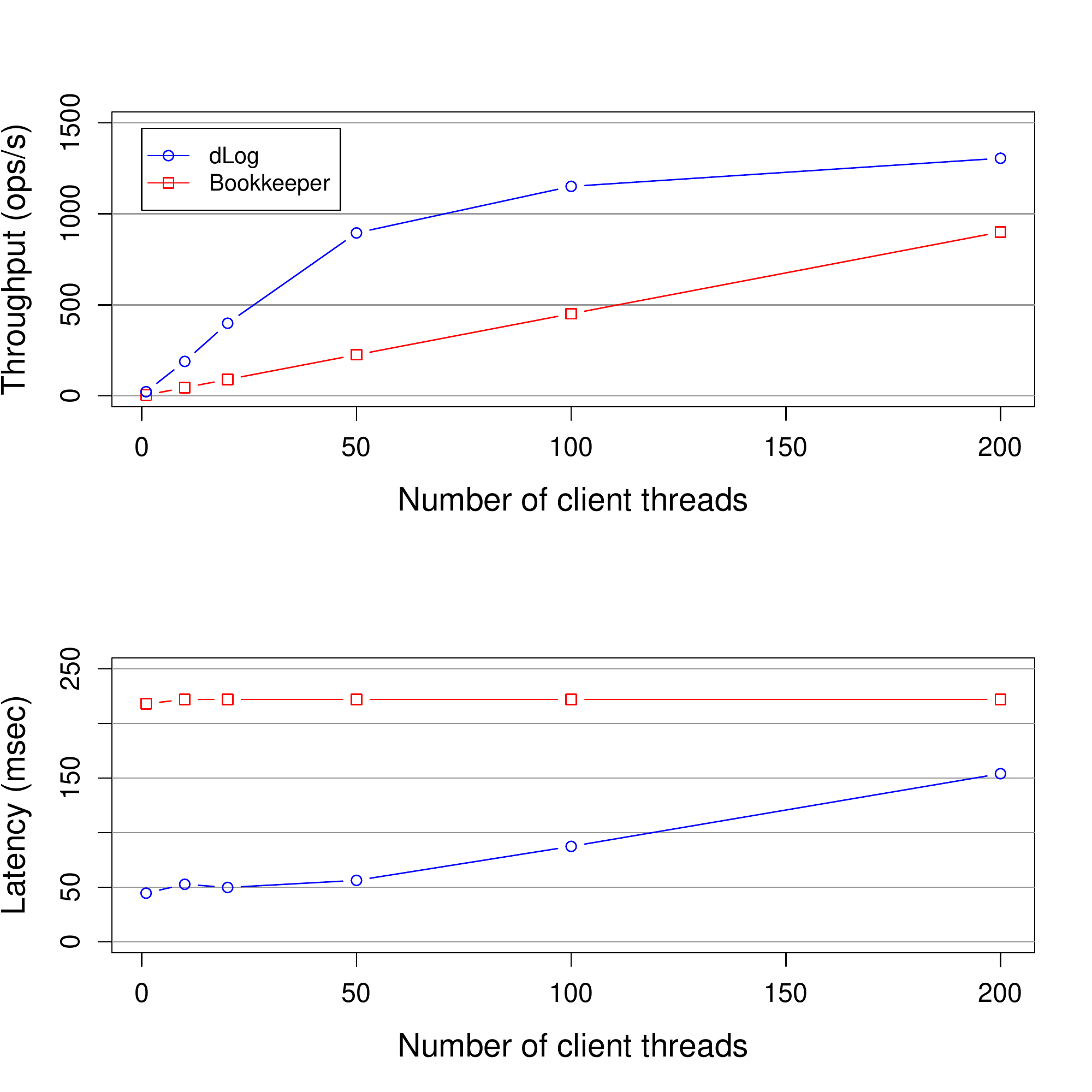} 
    \end{tabular}
\caption{Performance of dLog and Apache's Bookkeeper. The workload is composed of 1 Kbyte append requests. The graphs show throughput in operations per second (top) and average latency in msecs (bottom).}
    \label{fig:bookkeeper}
  \end{center}
\end{figure}

\textbf{Results.} Figure~\ref{fig:bookkeeper} compares the performance of our  dLog service with Apache Bookkeeper.
The dLog service consistently outperforms Bookkeeper, both in terms of higher throughput and lower latency.
With 200 clients, dLog approaches the limits of the disk to perform writes synchronously.
The large latency in Bookkeeper is explained by its aggressive batching mechanism, which attempts to maximize disk use by writing in large chunks.
%

\subsection{Scalability}
\label{ssec:scale}

In this section, we perform a set of experiments to assess the scalability of our proposed services.
We consider vertical scalability with dLog (i.e., variations in performance when increasing the resources per machine in a static set of machines) and horizontal scalability with MRP-Store (i.e., variations in performance when increasing the number of machines).

\subsubsection{Vertical scalability}
\label{ssec:dlogscale}

\textbf{Setup.} 
In this experiment, we evaluate vertical scalability with the dLog service by varying the number of multicast groups (rings). 
Each multicast group (ring) is composed of three processes, one of which assumes the learner's role only and the others are both acceptors and proposers. 
We perform experiments with up to 5 disks per acceptor, where each ring is associated with a different disk. 
Therefore, by increasing the number of rings, we add additional resources to the acceptors.
In each experiment, learners subscribe to $k$ rings and to a common ring shared by all learners, where $k$ varies according to the number of disks used in the experiment.
Processes in the rings are co-located on three physical machines. 
Clients are located on a separate machine and generate 1 KByte requests, which are batched into 32 KByte packets by a proxy before being submitted to Multi-Ring Paxos. 
The workload is composed of append requests only. 
Throughput is shown per ring. 
The reported latency is the average over all the rings. 

\begin{figure}[t]
  \begin{center}
    \begin{tabular}{c}  
      \includegraphics[width=\columnwidth]{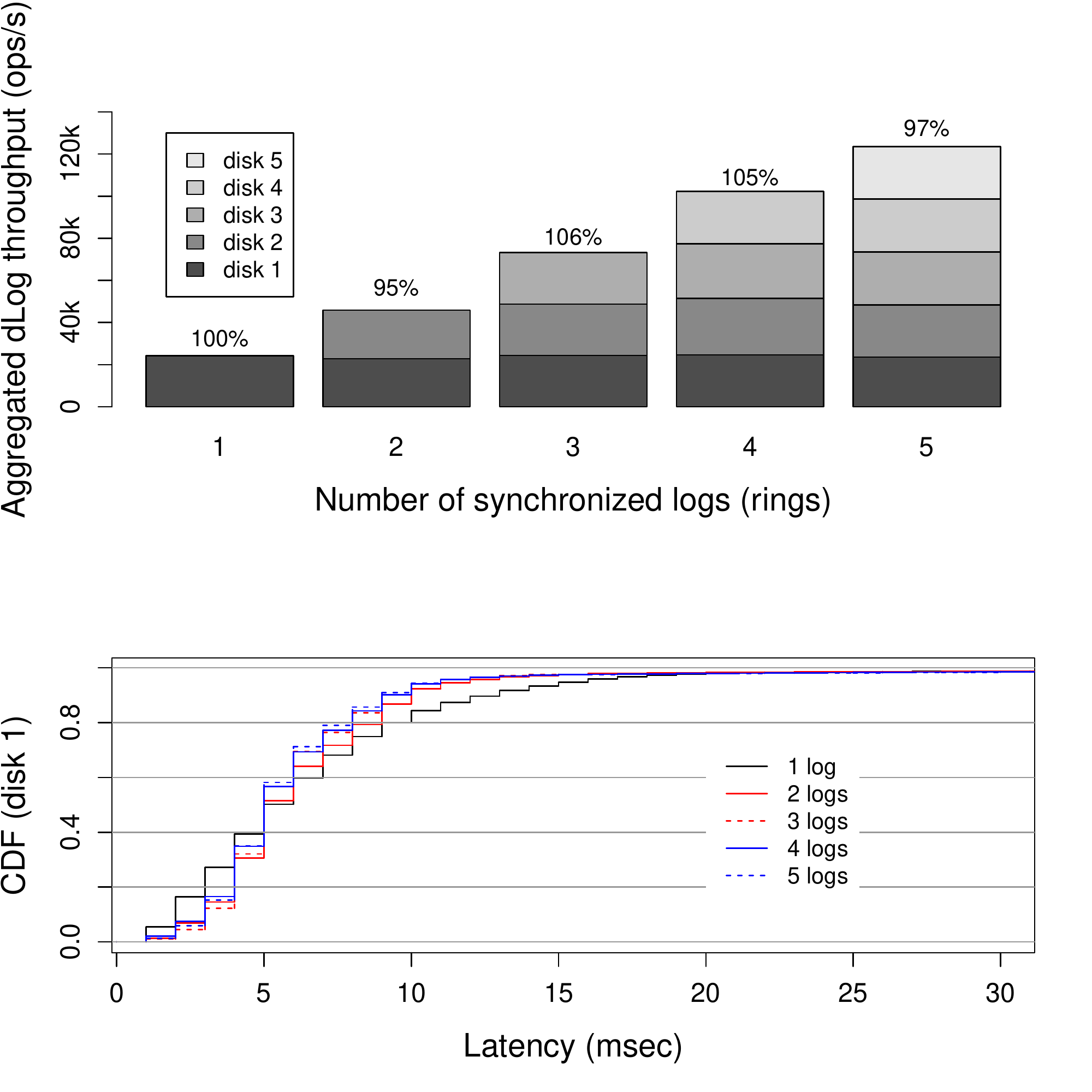} 
    \end{tabular}
\caption{Vertical scalability of dLog in asynchronous mode. The graphs show aggregate throughput in operations per second (top), and latency CDF in msecs (bottom).}
    \label{fig:scale-dlog}
  \end{center}
\end{figure}

\textbf{Results.} 
Figure~\ref{fig:scale-dlog} shows the throughput and latency of Multi-Ring Paxos as the number of rings increases. 
Throughput improves steadily with the number of rings.
The percent numbers shows the linear scalability relative to the previous values. 
The latency CDF corresponds to the reported throughput for writes to disk 1. 

\subsubsection{Horizontal scalability}
\label{ssec:mrpscale}

\textbf{Setup.} 
In this experiment, we evaluate horizontal scalability with the MRP-Store service, globally deployed across four Amazon EC2 regions (one in eu-west, two in us-west, and one in us-east). 
In each region there is one ring composed of a replica with three proposers/acceptors, and one client running on a separate machine. 
Replicas from all the rings are also part of a global ring. 
Clients send 1 KByte commands to their local partitions (rings) only. 
Each client machine batches the requests into packets of 32 Kbytes before sending them. 
The workload is composed of update requests only. 
Latency is measured in the us-west-2 region. 

\begin{figure}[t]
  \begin{center}
    \begin{tabular}{c}  
      \includegraphics[width=\columnwidth]{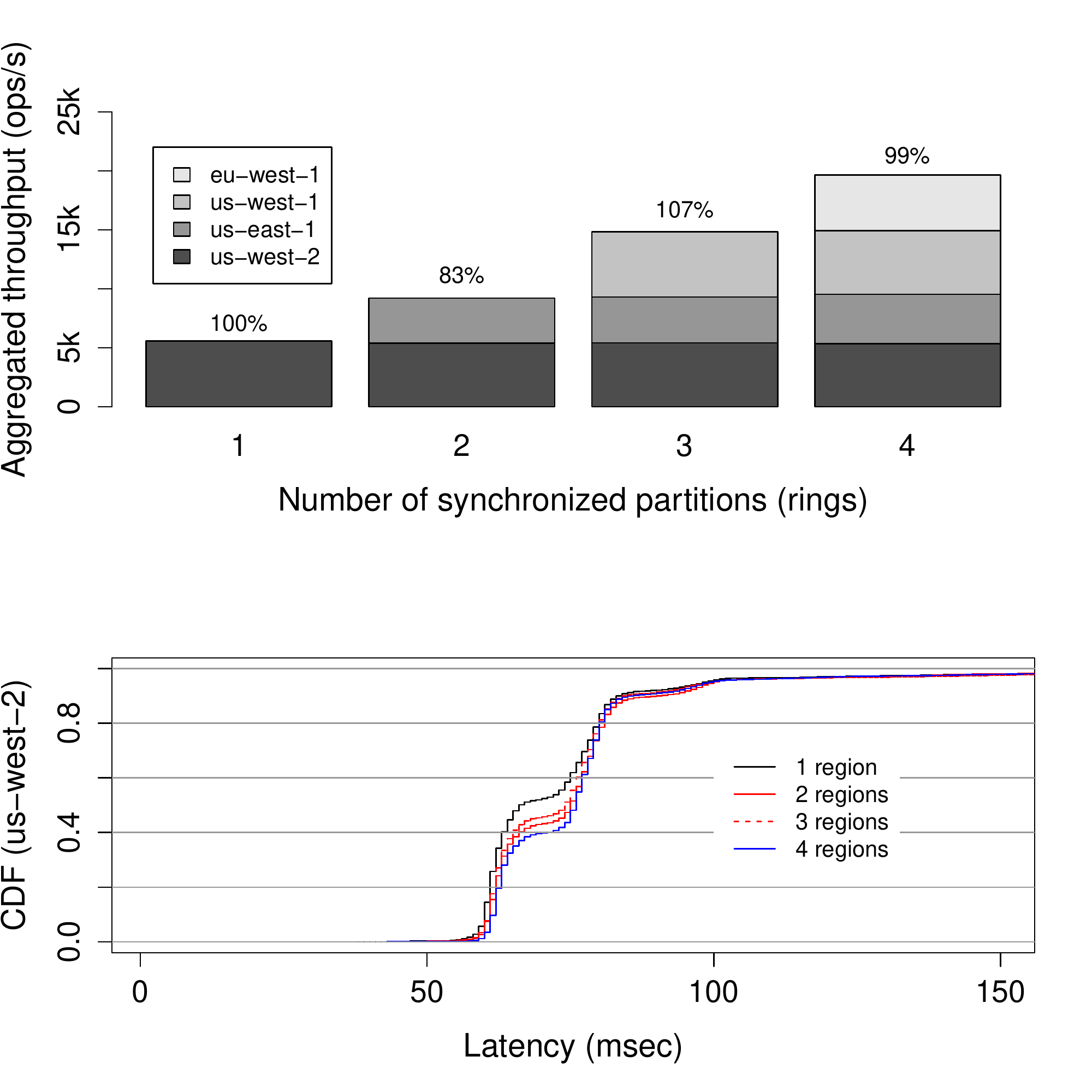} 
    \end{tabular}
\caption{Horizontal scalability of MRP-Store in asynchronous mode. The graphs show aggregate throughput in operations per second (top) and latency CDF in msecs in us-west-1 (bottom).}
    \label{fig:scale-mrp}
  \end{center}
\end{figure}

\textbf{Results.} 
Similarly to the dLog service, throughput increases as new partitions are added to the collection (see Figure~\ref{fig:scale-mrp}). 
As expected, latency is almost constant with the number of rings. 
We note that the local throughput of a region is not influenced by other regions, the reason for the scalability of the service. 
The percent numbers shows the linear scalability relative to the previous values. 

\subsection{Impact of recovery on performance}
\label{ssec:recovery-experiment}

In this section, we evaluate the impact of failure recovery on the system's performance using the MRP-Store service. 

\pagebreak
\textbf{Setup.} We deploy one ring with three acceptors, all performing asynchronous disk writes, and three replicas.
The system operates at 75\% of its peak load and there is one client generating requests against the replicas. 
The replicas periodically checkpoint their in-memory data store synchronously to disk to allow the acceptors to trim their log. 
One replica is terminated after 20 seconds and restarts after 240 seconds, at which point it retrieves the most recent checkpoint from an operational replica. 
The instances that are not included in the checkpoint will be retrieved directly from the acceptors.

\begin{figure}[t]
  \begin{center}
    \begin{tabular}{c}      
      \includegraphics[width=\columnwidth]{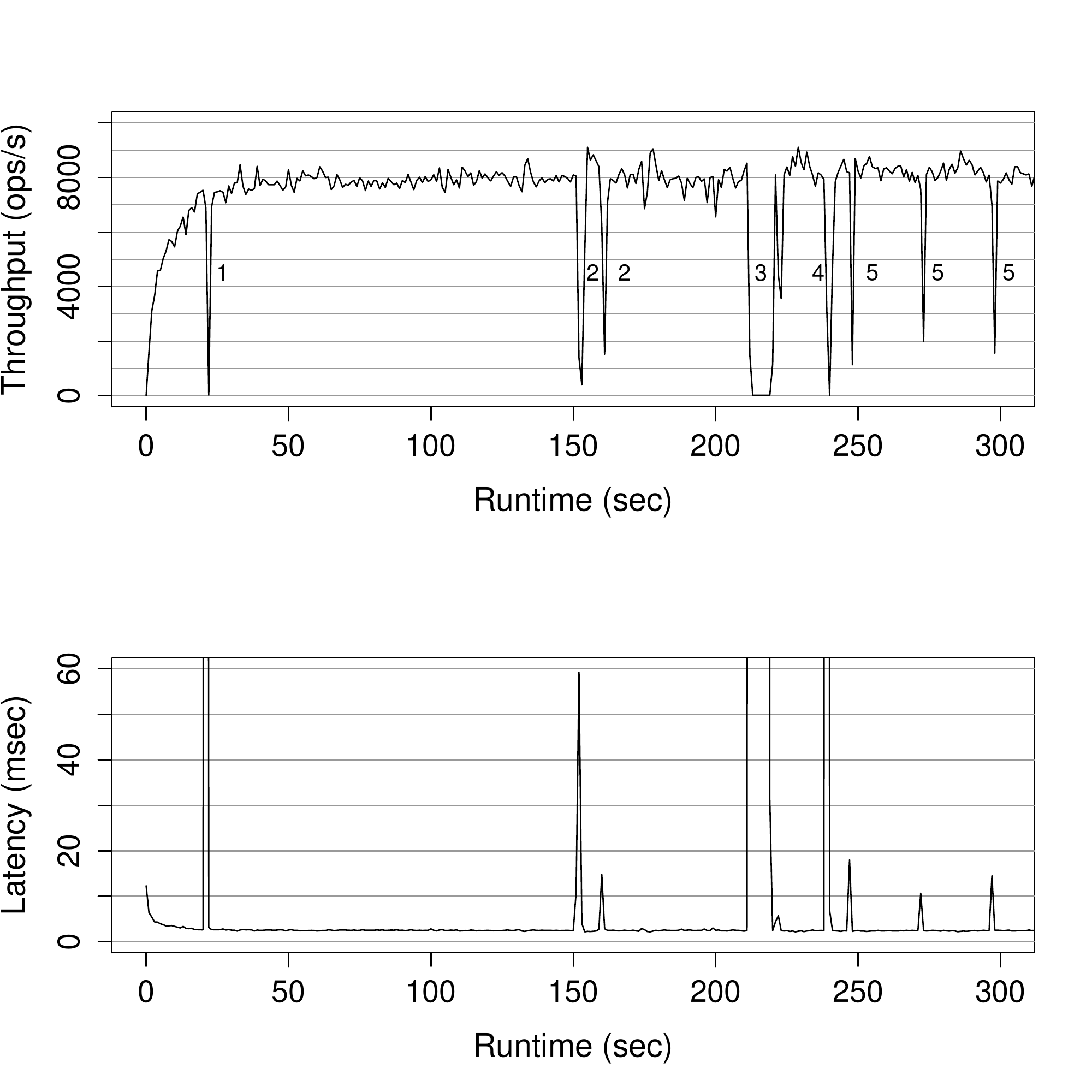} 
    \end{tabular}
\caption{Impact of recovery on performance (1: one replica is terminated; 2: replica checkpoint; 3: acceptor log trimming; 4: replica recovery; 5: re-proposals due to recovery traffic).}
    \label{fig:recovery}
  \end{center}
\end{figure}

\textbf{Results.} Figure~\ref{fig:recovery} shows the impact of recovery on performance. 
As seen in the graph, re-starting a terminated replica causes a short reduction in performance. 
Writing checkpoints synchronously to the disk does not disrupt the service either. 
We note that checkpoints are not written to disk at the same time by all the replicas and that the client waits only for the first response form any replica.  
Performance is mostly affected by trimming the acceptor logs and also when the recovering replica retrieves and installs a checkpoint. 

\section{Related work}
\label{sec:rwork}

In this section, we review related work on atomic multicast, distributed logging, and recovery.

\textbf{Atomic multicast.} 
The first atomic multicast protocol can be traced back to~\cite{BJ87b}, where an algorithm was devised for failure-free scenarios. 
To decide on the final timestamp of a message, each process in the set of message addressees locally chooses a timestamp, exchanges its chosen timestamps, deterministically agrees on one of them, and delivers messages according to the message's final timestamp. 
As only the destinations of a message are involved in finalizing the message's timestamp, this algorithm is scalable.
Moreover, several works have extended this algorithm to tolerate failures~\cite{fritzke1998amcast, GS01b, rodrigues1998scalatom, schiper2008inherent}, where the main idea is to replace failure-prone processes by fault-tolerant disjoint groups of processes, each group implementing the algorithm by means of state-machine replication. 
The algorithm in~\cite{DGF00} proposes to daisy-chain the set of destination groups of a message according to the unique group ids. 
The first group runs consensus to decide on the delivery of the message and then hands it over to the next group, and so on. 
Thus, the latency of a message depends on the number of destination groups.

While most works on multicast algorithms have a theoretical focus, Spread~\cite{ADM+04} implements a highly configurable group communication system, which supports the abstraction of process groups. 
Spread orders messages by the means of interconnected daemons that handle the communication in the system. 
Processes connect to a daemon to multicast and deliver messages.
To the best of our knowledge, Multi-Ring Paxos is the first high-performance atomic multicast library available for download.
Similarly to Mencius~\cite{Mencius}, coordinators in Multi-Ring Paxos account for load imbalances by proposing null values in consensus instances. 
Differently from Mencius, which is an atomic broadcast protocol, Multi-Ring Paxos implements atomic multicast by means of the abstraction of groups.
Multi-Ring Paxos's deterministic merge strategy is similar to the work proposed in \cite{aguilera2000detmerge}, which totally orders message streams in a widely distributed publish-subscribe system. 

\textbf{Distributed logging.} 
Atomic broadcast is not the only solution to totally order requests in a distributed environment. 
Distributed logging is an alternative approach, where appending a log entry corresponds to executing a consensus instance in an atomic broadcast protocol. 
CORFU~\cite{malkhi2012paxos} implements a distributed log with a cluster of network-connected flash devices, where the log entries are partitioned among the flash units. 
Each log entry is then made fault-tolerant using chain replication and a set of flash devices. 
New data is always appended to the end of the distributed log. 
To append a message, a client of CORFU (e.g., application server) retrieves and reserves the current tail of the distributed log through a sequencer node.
 Although appends are directly applied to the flash devices, the scalability of retrieving the log's next available offset is determined by the centralized sequencer's capacity. 
 In our dLog service, the increasing append load is smoothly absorbed by adding new rings to the ensemble, and is not subject to central components. 
 Disk Paxos~\cite{gafni2003disk} is another implementation of a distributed log that does not rely on a sequencer. However, Disk Paxos is not network efficient since for appending new data clients always contend over the log entries. An advantage with CORFU and similar systems~\cite{hartman1995zebra} is that the distribution of appends among the storage units can be balanced. Tango~\cite{balakrishnan2013tango}, builds on CORFU to implement partitioned services, where a collection of log entries is allocated to each partition. 
The replicas at each partition only execute the subset of the log entries corresponding to their partitions, and skip the rest. Globally ordering the entire set of log entries simplifies ensuring consistency with cross partition queries. 
However, the number of partitions a service can be divided into is limited by the log's capacity at handling the appends. 
In our dLog service, an unbounded number of partitions can be created by adding new rings; moreover, queries concerning disjoint partitions are not globally ordered. 



\textbf{Recovery.} Recovery protocols often negatively affect a system's performance. Several optimizations can be applied to the logging, checkpointing, and state transfer to minimize the overhead of recovery as we discuss next. 

\emph{Optimized logging.} A common approach to efficient logging is to log requests in batches~\cite{bessani2013efficiency,clement2009making,castro1999practical,singh2009zeno,kotla2007zyzzyva}.
Since stable storage devices are often block-based it is more efficient to write a batch of requests into one block rather than to write multiple requests on many different blocks. Another optimization is to parallelize the logging of batches~\cite{bessani2013efficiency}. Parallel logging benefits most the applications in which the time for processing a batch of requests is higher than the time required for logging a batch. The overhead of logging can be further reduced by using solid-state disks (SSD) or raw flash devices instead of magnetic disks~\cite{rao2011using}. Similarly, in our dLog service we support both harddisks and SSDs, and synchronous and asynchronous disk writes to enable batched flushes to the disk. 

\emph{Optimized checkpointing.} Checkpoints are often produced during the normal operation of a system, while processing of the requests is halted~\cite{castro1999practical,lamport1998part,singh2009zeno,kotla2007zyzzyva,rao2011using}. 
Not handling requests during these periods makes the system unavailable to clients and reduces performance. 
If instead processes take checkpoints at non-overlapping intervals, there will always be operational processes that can continually serve the clients. 
Building on this idea, in~\cite{bessani2013efficiency} processes schedule their checkpoints for different intervals. As the operation of a quorum of processes is sufficient for their system to make progress, a minority of processes can perform checkpointing while the others continue to operate. 
Another optimization is to use a \emph{helper} process to take checkpoints asynchronously~\cite{clement2009upright}. In this scheme, two threads, primary and the helper, execute concurrently. While the primary processes requests, the helper takes checkpoints periodically. Similarly, in our dLog service replicas can take snapshots at different non-overlapping intervals.


\emph{Optimized state transfer.} State transfer has its own implications on performance. 
During state transfer, a fraction of the source processes's resources (e.g., CPU, network) are devoted to the transmission of the state, which is not to the advantage of performance. 
To protect performance, state transfer can be delayed to a moment in which the demand on the system is low enough that both the execution of new requests and the transfer of the state can be handled~\cite{hunt2010zookeeper}. 
Another optimization is to reduce the amount of transferred information. Representing the state through efficient data structures~\cite{castro1999practical}, using incremental checkpoints~\cite{clement2009upright,castro:2003base}, or compressing the state are among these techniques. In~\cite{bessani2013efficiency}, authors propose a collaborative state transfer protocol to evenly distribute the transfer load across replicas. 






\pagebreak
\section{Conclusions}
\label{sec:conclusions}

When replicating services in large-scale settings, one common approach to scale performance and reduce latency is to weaken consistency.
Weak consistency, however, places the burden on the service users, who must cope with non-intuitive service behavior.
Providing strong consistency in globally distributed settings requires ordering requests across multiple datacenters. 
While some proposals impose a total order on requests, some other systems partially order requests across datacenters by means of ad hoc protocols (e.g., two-phase commit).
In this paper, we argued that atomic multicast is the proper abstraction to implement highly available and scalable systems without sacrificing consistency. 
%
We showed the practicality of our argument by implementing a high-performance atomic multicast library equipped with efficient recovery to build globally distributed, consistent, and durable key-value store and logging services. 
Moreover, the results of our experiments demonstrate both horizontal and vertical scalability of our proposed techniques. 


\bibliographystyle{abbrv}
\bibliography{main}

\end{document}